\documentclass{article}

\usepackage{PRIMEarxiv}

\usepackage[utf8]{inputenc} 
\usepackage[T1]{fontenc}    
\usepackage{hyperref}       
\usepackage{url}            
\usepackage{booktabs}       
\usepackage{amsfonts}       
\usepackage{nicefrac}       
\usepackage{microtype}      
\usepackage{lipsum}
\usepackage{fancyhdr}       
\usepackage{graphicx}       
\graphicspath{{media/}}     
\usepackage{multirow}
\usepackage{multicol}
\title{ALGORITHM FOR AGC INDEX MANAGEMENT AGAINST CROWDED RADIO ENVIRONMENT
}

\author{
  Morgane Joly, Fabian Rivière \\
  NXP Semiconductor\\
  Caen, France\\
  \texttt{\{$morgane.joly\_1, fabian.riviere$\}@nxp.com} \\
   \And
  \'Eric Renault\\
  LIGM, Univ. Gustave Eiffel, CNRS, ESIEE Paris \\
  Noisy-le-Grand, France\\
  \texttt{$eric.renault@esiee.fr$} \\
}

\begin{document}
\maketitle

\begin{abstract}
This paper describes a receiver that uses an innovative method to predict, according to history of receiver operating metrics (packet lost/well received …), the optimum automatic gain control (AGC) index or most appropriate variable gain range to be used for next packet reception, anticipating an interferer appearing during the payload reception. This allows the receiver to have higher immunity to interferers even if they occur during the gain frozen payload reception period whilst still ensuring an optimum sensitivity level.
As a result, the method allows setting the receiver gain to get an optimum trade-off between reception sensitivity and random interferer immunity.

\end{abstract}

\keywords{ML, BLE, Countermeasure, CR}

\section{Introduction}
Connected devices are part of everyday life. The proliferation of connected portable devices such as mobile phones, laptop, smart watches, tablets, or non-portable connected devices such as TV, video game console saturates the environment with RF signals. 
In parallel to the reception of desired data from its communication partner(s), such connected devices receive also unwanted signals, so called interferers. 
The interferers, especially from Wi-Fi signals, can occur in a random manner in the form of a signal burst of variable duration and have a signal strength possibly much higher than the desired signal. Interferers with a high signal strength can cause saturation of the receiver preventing proper reception of the desired data. 
Some techniques tackle this issue by continuously monitoring the received signal strength and adjust immediately the receiver gain to avoid saturation whilst still maintaining the highest sensitivity level.

\begin{figure}[!htt]
 \centering
      \includegraphics[scale=0.62]{ 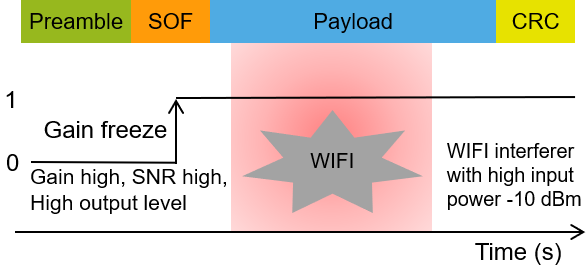}

\caption{IEEE 802.15.4 packet reception corruption Wi-Fi packet interferer occurring during payload reception period}
        \label{fig:packet_interf}
\end{figure}

However, when operating popular wireless communication protocols such as Wireless PAN (Bluetooth, BLE, Zigbee…), the receiver is not allowed to adjust the gain during the data payload. RF receivers for these communication protocols adjust then the gain during a time interval prior to the payload reception based on the real-time received signal and freeze the gain just before switching to the payload reception period. This is illustrated in figure~\ref{fig:packet_interf}. Due to the random nature in occurrence and strength level, interferers may appear during the data payload, receiver may saturate causing data loss.

\newpage

This paper describes an innovative method to predict, according to history of receiver operating metrics (packet lost/well received …), the optimum AGC index or most appropriate variable gain range to be used for next packet reception, anticipating an interferer appearing during the payload reception.
This allows the receiver to have higher immunity to interferers even if they occur during the gain frozen payload reception period whilst still ensuring an optimum sensitivity level. Next slide illustrates benefits of proper AGC setting.

\section{Related work}

\label{sec:related}

Integration of artificial intelligence in the domain of the non-collaborative coexistence of wireless protocol is a well-documented subject. The most common ways to improve the quality of service of radio frequency receiver are divided into three categories~:

\subsection{Intelligent channels assignment}
The intelligent channels assignment search to improve the Adaptative Frequency Hopping with usually a channels whitelisting or blacklisting. ML algorithm can be used for this task as proposed in~\cite{ML_comparison} which compares GPOMDP, Episodic-Reinforcement, and True Policy Gradient to enhance channels assignment. Giral and al.~\cite{115080} proposed a deep-learning-based algorithm to spectral decision in cognitive radio networks. They noticeably use the real behaviour of users to collect their data to translate them into RGB value to train a visual pattern recognition DL-algorithm.

To the best of our knowledge, Nikoukar and al.~\cite{9217227} are the only team to propose a model (LSTM-based) that address the specific subject of Wi-Fi interference prediction on BLE channels.

\subsection{Interference and intrusion detection}
The detection of interference and intrusion contribute to better detect packets and allows the reconstruction of packets degraded by malicious activity surrounding device.

O'Mahony and al.~\cite{9221332} detect jamming in global positioning system (GPS) signals thanks to a random forest and pre-processed data. This team studies also in~\cite{9180209} the Random Forest-based detection of malicious interference within a Zigbee communication using In-phase (I) and Quadrature-phase (Q) samples. As T. Kikuzuki and al.~\cite{8108218} which propose an algorithm to identify nearby protocol to improve detection sensitivities while there is emission overlap thanks to a new layout of signal decoding. Barac and al.~\cite{6827630} designs a classifier with a specifically designed packet to detect the presence of an harmful interferer. Grimaldi \cite{8570750} proposes a similar system of detection and identification of interference using supervised learning. Lee and al.~\cite{denoise} produce a new kind of denoiser based on neural network that learns from the ambient noise to subtract it to the signal, this kind of system can be used for signal restoration and improve symbol demodulation. The study~\cite{5962637} differentiate the events that may have caused the loss of the packet and adapt the emission of packets to the spectral environment.

Internal receiver metrics to detect Wi-Fi emission have been interestingly used in~\cite{8305459} which have developed a Wi-Fi scanning manager application to enhance the delay and reduce the energy consuming in the discovering available access points phase. 

\subsection{Spectrum prediction and nearby protocol detection}
Natively, devices cannot detect nearby protocol to apply the best counter-measure associated to them.
In the study~\cite{9221276}, Sudharsan et al. proposes to add flexibility in the choice of the protocol used by the radio with a Support Vector Regression (SVR) that predicts the RSSI evolution to guide the choice of the best protocol in the future environmental conditions. Nevertheless, it implies integrating in the hardware all the possible protocols to allow a device to have the best choice panel.
Wang and Zhang in \cite{7852914} use a decision tree to take in consideration the MAC layer in the spectrum sensing scheme to design a new strategy of emission and improve the quality of service of secondary users by wisely choose the instant of emission to avoid primary user interference. 

\subsection{AGC improvement}
To the author knowledge, no improvements proposition of the AGC algorithm by ML had been proposed yet but the use of others statistical tools of prediction can be found in the patent database showing the interest shown by the scientific community over the years in improving this part of radio. 

The applications \cite{patent1}, \cite{patent2}, \cite{patent3} use classical engineering methods for gain adjustment decision. 

Wang and al. propose to apply voice detection by detecting the number of occurrences of the peak value during a second predetermined time segment within the measurement window. The media processing system identify the speech portion by using  classical engineering methods for gain adjustment decision (peak detection and detection threshold use to adjusts a gain). 

Rothaar and al. use a statistical (averaging and watching for correlations) model to determine the pulse rate and pulse width of a radar and predict the timing of the next pulse. The AGC can raise or lower gain levels for the duration of the pulse; or ignore the pulse, if it is of a very short duration, so that the AGC level will coast through the event. 

Li and al.  combines three concepts, a statistical database, determination of evolution AGC trend (increase of AGC, decrease, uncertain), and threshold (low and high threshold and mean value). Thus, they method will increase, decrease or retain unchanged the AGC for the next slot.
The method comprises predicting an AGC setting to be used based on statistical data with respect to a plurality of previously stored AGC settings.

\cite{patent4}, \cite{patent5} disclose a method to instantaneously adjust the receiver gain based on the strength of the received RF signal to ensure the highest sensitivity while preventing receiver’s saturation. Such methods are adapted to optimize the sensitivity to the highest possible level but cannot prevent saturation due to interferer occurring during the data frame causing data lost.

\cite{patent6} propose a method with improved interferer handling. It teaches a receiver with gain adjustment during the preamble. The gain adjustment is done based on the desired data strength (in-band signal) an includes a predetermined margin to avoid saturation in case an interferer occurs during the data frame where the gain is frozen.

\section{Description}
\label{sec:description}
This paper describes a method of monitoring and controlling the AGC gains of the receiver to improve coexistence of BLE/15.4 radio subject to an interferer. A multi-variables (RSSI, CRC...) statistical tool is used based on a Machine-Learning algorithm to predict the AGC gain index to be used in the next reception slot. This prediction allows the anticipation of a late arrival (after the AGC index freeze) of an interferer that may jeopardise the receiver gain index natively computed by AGC Hardware State Machine (before AGC index freeze). This prediction allows the identification of an unsustainable environment and recommends adapted countermeasures such as adapting the AGC gain index range according to the previously detected interferences and/or channel hopping table adjustment.

This prediction is done before the radio goes in sleep mode. No extra radio activation time slot requested. Hence, when the radio wakes up then AGC index (upper limit) to use is already available. Its value is optimum to adapt to interferer fluctuant characteristics. Proper gain setting reduces packet loss and consequently packet retries. This improves the PER (Packet Error Rate) and energy consumption.

\section{Improvement by proper AGC gain selection – Proof of concept}

\begin{figure}[!ht]
 \centering
      \includegraphics[scale=0.72]{ 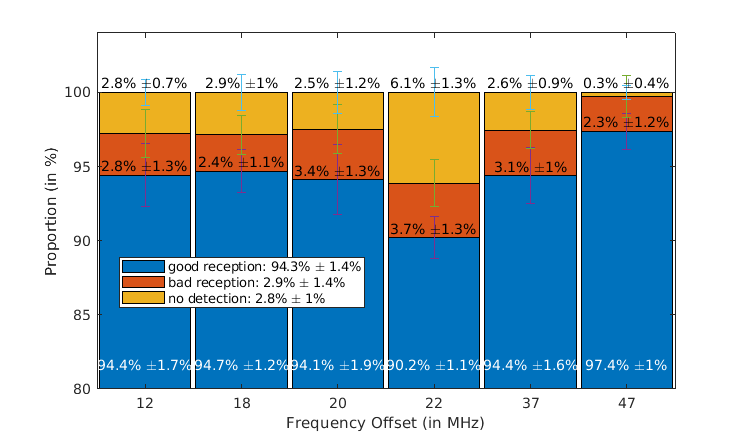}
    	\caption{Benefits of proper AGC setting to get receiver higher immunity to interferers}
        \label{fig:improv_proof}
\end{figure}

Figure~\ref{fig:improv_proof} illustrates the benefits of proper AGC setting to get receiver higher immunity to interferers even if they occur during the gain frozen payload reception period. This paper proposes an innovative method to predict this proper AGC setting. To get this illustration we studied packets reception behavior for two test cases:
\begin{itemize}
    \item The interferer arrives before the AGC index freeze
    \item The interferer arrives after the AGC index freeze
\end{itemize}
We have selected a set of specific wanted packet / interferer configuration (relative level / frequency offset) where we could observe an evolution of reception status, namely a packet which is well received when the interferer arrives before the AGC index freeze, but badly received when the interferer arrives after AGC freeze.
By reemitting the wanted packets with an interferer arriving after the index freeze and by bypassing the AGC index with the value determined when the interferer already present before wanted packet arrival, we observe a good amount of evolution, i.e. $61\%$ of the packet collection studied evolve toward a good reception.
Note the difference between a native AGC determined gain index obtained with interferer present before and after AGC freeze, as this is used as an error signal to train ML algorithm, which can then predict an optimum AGC index.

The table below shows the packet population for each frequency offset studied. 

\begin{table}[h]
\centering 
\caption{Packet reception status classes}
\begin{tabular}{|c||c|c|c|c|c|c|}
\hline
    MHz offset  & 12& 18 &20& 22 &37& 47     \\
   \hline
   Number of packet analysed & 13 &9& 15& 15& 14& 20   \\
   \hline

\end{tabular}
\end{table}

\newpage
\section{Paper organisation}
The subsequent description consists of the following sections:
\begin{itemize}
    \item Receiver block diagram
    \item ML algorithm training method
    \begin{itemize}
        \item Flowchart of packet processing and ML training/test/validation
 \item Receiver status definition
 \item Class definition for optimal AGC gain index calculation
 \item Individual packets dataset definition
 \item Signal dataset definition with bursted interferer profile 
 \item Training / test sub-dataset definition
 \item Pre-processing of previous sub-datasets with sliding window 
 \item ML generation model by training sequence

\end{itemize}
\item Receiver block diagram
\item ML algorithm training method
\begin{itemize}

 \item Flowchart of packet processing and ML training/test/validation
 \item Receiver status definition
 \item Class definition for optimal AGC gain index calculation
 \item Individual packets dataset definition
 \item Signal dataset definition with bursted interferer profile 
 \item Training / test sub-dataset definition
 \item Pre-processing of previous sub-datasets with sliding window 
 \item ML generation model by training sequence
 \end{itemize}
\item ML aided AGC in functional mode
\begin{itemize}

 \item Flowchart of ML data processing for AGC gain selection 
 \item Metrics (data) value collection during packet reception / sliding window 
 \item Potential integration of the solution in radio (scenario 4)
 \item Experiment result : PER versus power blocker – interferer at 12 MHz offset – scenario 4
\end{itemize}
    \end{itemize}
\section{Receiver Block Diagram}
\begin{figure}[!ht]
 \centering
      \includegraphics[scale=0.62]{ 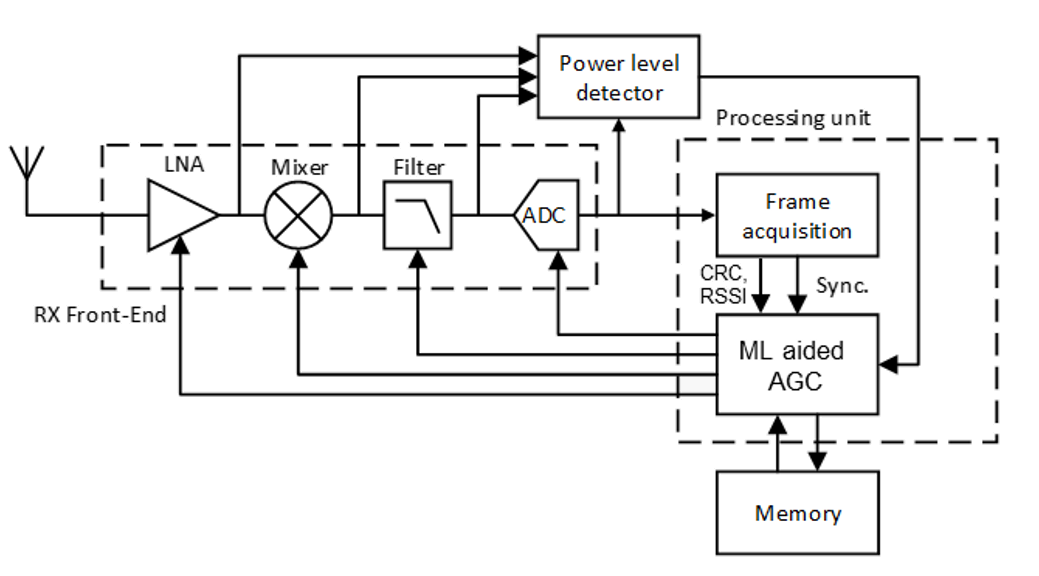}
    	\caption{Receiver with ML aided AGC block diagram}
        \label{fig:radio_chain}
\end{figure}

Figure~\ref{fig:radio_chain} shows a receiver consisting of an amplifier (LNA) connected to an antenna followed by a power level detector, a down converter, a low-pass filter, an analog-to-digital converter (ADC), a processing module and a memory block. 
The processing module includes a frame acquisition, an automatic gain controller (AGC) assisted by a Machine Learning (ML) Engine. 

In Figure~\ref{fig:improv_proof}, the power level detector is coupled only to the output of the LNA and the AGC adjusts only the gain of the LNA. In another embodiment, the power level detector is coupled to the output of a plurality of blocks (LNA, Mixer, LPF, ADC) and the AGC can adjust the gain of each individual block (LNA, Mixer, LPF, ADC) independently. 

The frame acquisition is coupled to the AGC, detects the different portions of the frame (i.e. preamble, start-of-frame, data payload…) and provides a frame synchronization signal to the AGC as well as features other metrics (i.e. CRC, SNR, RSSI...) to feed ML engine. The data frame can be formed according to wireless communication standards such as various categories of IEEE802.15.4, Bluetooth, BLE. Although the exemplary embodiment is based on IEEE802.15.4 data frame format, the proposed method can be easily adapted to other data frame formats like BLE. The ML aided AGC is coupled to the memory block to store and retrieve receiver metrics and well as to store ML learning database.

The ML engine is fed by receiver internal metrics during a sliding time window that will predict the allowed AGC gain range (upper limit) or best AGC gain setting based on gathered statistical data during this time window.

Various determination methods can be applied depending on the interferer types. In normal operation of the radio, it is assumed the ML model  has already been trained and ML model dataset is available.

\section{ML training process - Dataset creation}
\subsection{Flowchart of packet processing and ML training \& test \& validation}

\begin{figure}[!ht]
 \centering
      \includegraphics[scale=0.62]{ 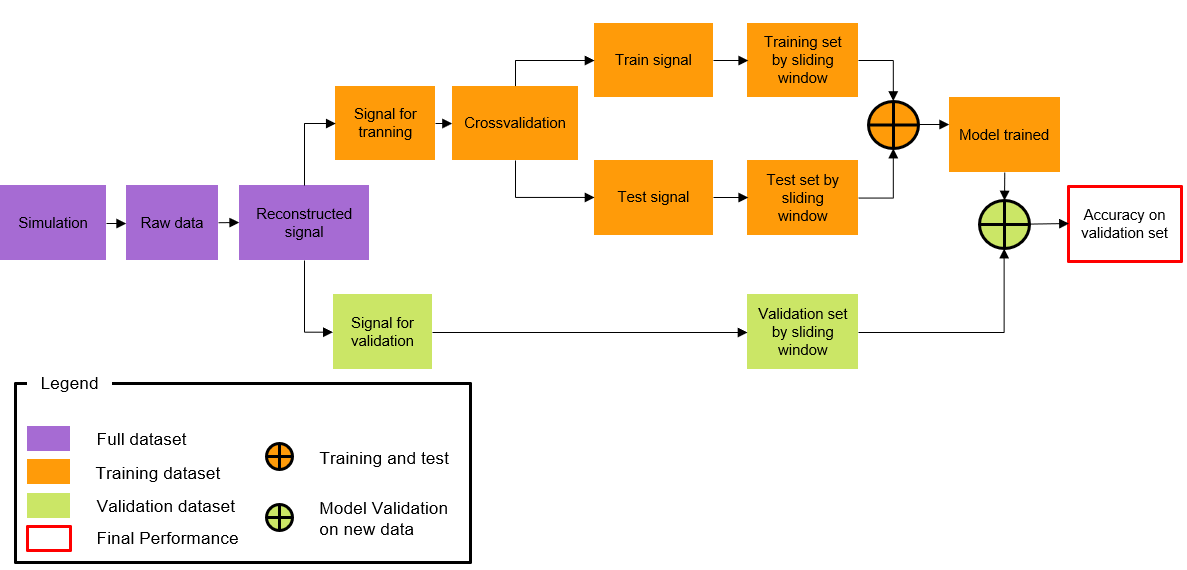}
    	\caption{Flowchart of packet processing and ML training/test/validation}
        \label{fig:data_flow}
\end{figure}

Figure~\ref{fig:data_flow} shows the data processing performed to train the ML engine and summarizes the training model steps. During this process, the ML engine will capture (characterize) the receiver metrics dataset behavior versus an extensive collection of wanted packet/interferer configuration (relative level / frequency offset / bursts sequences), driving the antenna port, to derive the optimum AGC gain index (upper limit) to be used for the next packet to receive.

\newpage

Hence, Figure~\ref{fig:data_flow} provides a view of the  radio metrics dataset (raw data obtained from frame acquisition or power level detectors functional blocks) treatment from collect to transformation in dataset of training and validation usable by machine learning (ML).

Three tools have been created:
\begin{itemize}
    \item Data collection
\item Reconstruction of signal by real-like ratio cycle of Wi-Fi
\item Cross-validation for signals
\end{itemize}

Next sections provide some details about treatment steps done in sequence.

\subsection{Receiver status definition}

The radio metrics dataset (raw data obtained from the frame acquisition functional block in this case) treatment is used to obtain a clear status of the reception of a packet for a given wanted packet / interferer configuration.
The reception status is the combination of two parameters got from the packet reception process.
\begin{itemize}

 \item Cyclic redundancy check (CRC)
 \item Access Address Found (AA Found)
 \end{itemize}
If no Access Address field has been detected during the radio active time, then no packet has been processed
\begin{itemize}
 \item CRC = 0 : normal working, no packet has been detected => no reception
 \item CRC = 1 : abnormal working, radio dysfunction
 \end{itemize}
If an Access Address has been detected during the radio active time => a packet has been received
\begin{itemize}
 \item CRC = 0 : no error during the reception => Good reception
 \item CRC = 1 : error during the reception => Bad reception
 \end{itemize}
The table below shows a summary of the reception status.

\begin{table}[h]
\centering 
\caption{Packet reception status classes}
\begin{tabular}{|c||p{15mm}|p{20mm}|}
\hline
     CRC/AAFound           & 0   &  1   \\
   \hline
   0            & No reception (NR)                 &  Bad reception    \\
   \hline
   1            & Radio Error               &  Good reception (GR)     \\
    \hline
\end{tabular}
\end{table}
\newpage

\subsection{Class definition for Optimal AGC calculation
}
\begin{figure}[!ht]
 \centering
      \includegraphics[scale=0.62]{ 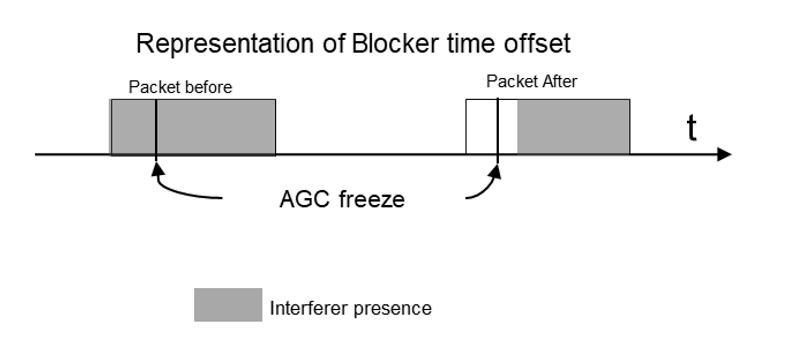}
    	\caption{Interferer arrival time versus AGC freeze event}
        \label{fig:inter_offset}
\end{figure}

The creation of the labelled class consists of comparing the reception status for a given wanted packet/interferer configuration with 2 different versions of the blocker arrival time illustrated in Figure~\ref{fig:inter_offset}.

\begin{itemize}
 \item A wanted packet is emitted in the same condition
 \item Same signal wanted power level
 \item Same wanted packet / interferer frequency offset
 \item Same blocker power level
 \item Same simulation seeds
 \item The only changing parameter is the time offset of interferer arrival time relative to wanted packet start
 \item Before the freeze of AGC $\Rightarrow$ Then native AGC selects $AGC_{before}$ gain index value at AGC freeze
 \item After the freeze of AGC $\Rightarrow$ Then native AGC selects $AGC_{after}$ gain index value at AGC freeze
 \end{itemize}

The optimal AGC ($AGC_{optim}$) gain index value is chosen according to the best reception status possible
\begin{itemize}
 \item If there is no good option (i.e. the packet is always bad received or cannot be detected when interferer arriving before the AGC freeze) then the $AGC_{optim}$ is set to the value “X”
\item Meaning the reception in the given channel is not sufficiently good for the radio
\item Several countermeasure options maybe 
\item A threshold can be set to blacklist the channel for a certain amount of time
\item The use of this channel is reduced
\end{itemize}
The table below shows a summary of the optimal AGC calculation :
\begin{figure}[!ht]
 \centering
      \includegraphics[scale=0.62]{ 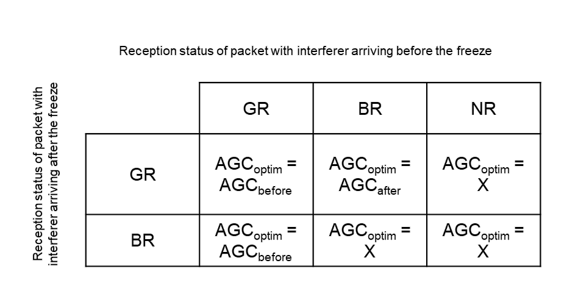}
    	\caption{Summary of the optimal AGC calculation}
            \label{tab:calcul_AGC}
\end{figure}

\subsection{Individual packet dataset definition}
The dataset is generated for wanted packet / interferer set of parameters swept at antenna port and in the simulation or in lab
\begin{itemize}
 \item Wanted power level
 \item Blocker power level
 \item Blocker versus Wanted packet Frequency offset 
 \end{itemize}
The radio metrics generated during the reception of packet are collected (RSSI, SNR, CRC, AA Found, AGC…)
\begin{itemize}
 \item Will be used as ML input features
 \item To create ML expected output then process described before is used
 \end{itemize}
At this step, the behaviour of the radio and wanted $ AGC_{optim}$ value is determined
\begin{itemize}
 \item Next step is to prepare a « synthetic signal structure” dataset that will take into account bursted nature of interferer
 \item Then sequences of wanted packet / interferer different configurations are then available for timing window concept usage
 \end{itemize}
The table below shows an example of features collected versus wanted packet/ interferer 
configuration sweep and the $ AGC_{optim}$ finding process.

\subsection{Wi-Fi specification - Signal Dataset definition with bursted interferer profile}

\begin{figure}[!ht]
 \centering
      \includegraphics[scale=0.50]{ 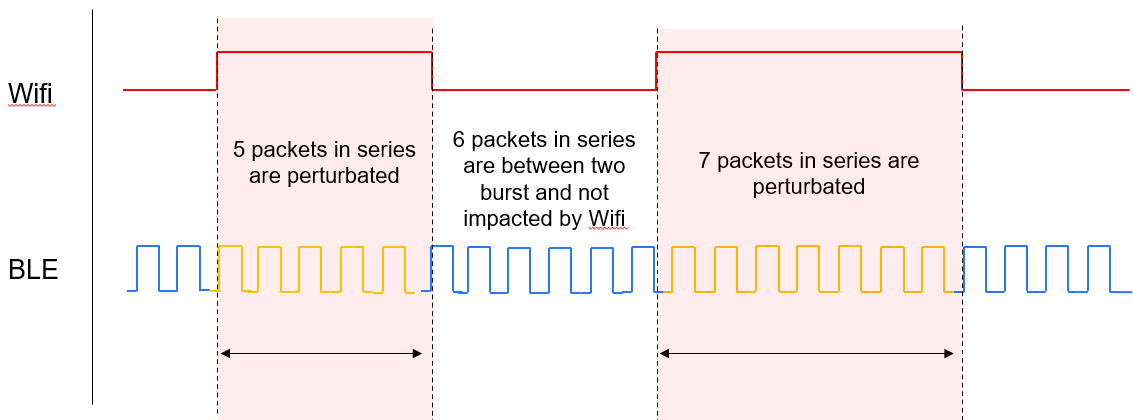}
    	\caption{ Example of Wi-Fi pattern scenario that disturbs BLE signal}
        \label{fig:serie_packet}
\end{figure}

Wi-Fi is a bursted signal which is emitted according the user activity~\cite{anatWifi}, different cycling ratio can be observed.
To create realistic stress of a BLE communication perturbated by a Wi-Fi, Individual packet dataset previously determined is used
\begin{itemize}
 \item First a reference wanted signal power is chosen
 \item Then interferer power level range is chosen according to the chosen pattern
 \begin{itemize}
\item High : [ -23;0] dBm
\item Mean : [-46; -23] dBm
\item Weak : [-70;-47]dBm
\item Absent : < -71 dBm
\end{itemize}
 \item Finally, wanted packets are sequenced randomly in the subset created
 \begin{itemize}
\item This sequence defines “synthetic signal structure”
\end{itemize}
\end{itemize}
An example sequence is shown in Figure~\ref{fig:serie_packet}. The alternation of packet group represent signal cyclic ratio. Each Wi-Fi scenario has a typical pattern that disturbs the BLE signal.

\newpage

\subsection{Training/Test sub-dataset definition}
\begin{figure}[!ht]
 \centering
      \includegraphics[scale=0.70]{ 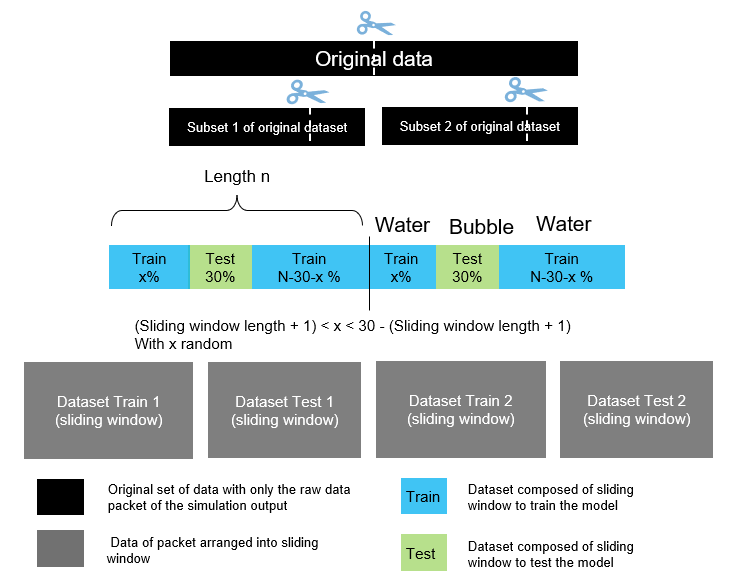}
    	\caption{Training/Test sub-dataset definition}
        \label{fig:crossvalid}
\end{figure}

We assume now that past activity of the receiver will be observed over a sliding window of N packets duration. We need to create the subset of ML train and test which will be used to predict the optimal AGC gain index upper limit ($AGC_{optim}$) to be used at N+1 packet reception start period, while ensuring the reduction of the overfitting. To do so, the cross-validation concept is adapted to the synthetic signal structure while ensuring the balanced with the data split of 70\% in the train and 30\% in the test. The previous signal dataset is cut into as many folds as necessary, then a continuous portion of 30\% of the fold is randomly selected for test subset, the other 70 \% part is for train subset. All the respective part are regrouped, and finally the sliding window is applied. The ML model is trained from scratch with the same dataset but different train and test subset to ensure the validity and the stability of the results. The process is shown in Figure\ref{fig:crossvalid}.

\newpage

\subsection{Pre-processing of previous subset with sliding window}
\begin{figure}[!ht]
 \centering
      \includegraphics[scale=0.70]{ 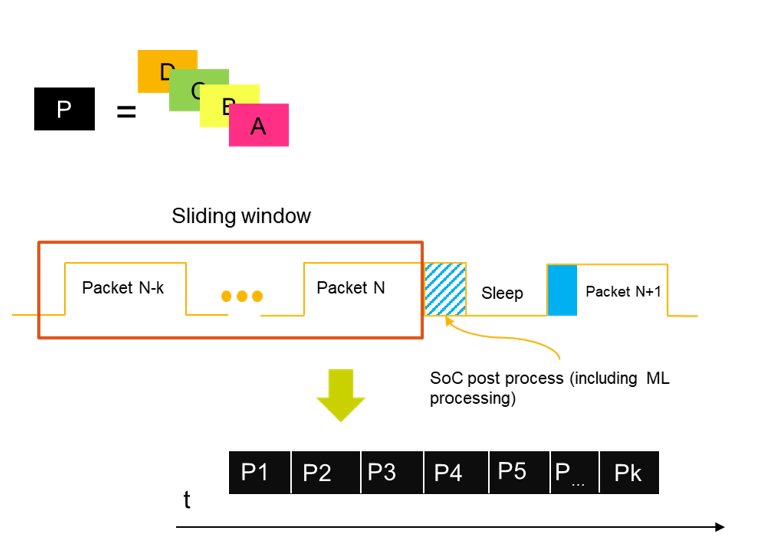}
    	\caption{Pre-processing of previous subset with sliding window}
        \label{fig:packet_in_ML}
\end{figure}

The sliding window (Figure~\ref{fig:packet_in_ML}) pre-processes the data packet before to pass though the model This step gathers all the data of each signals from internal metrics of each packets and the $AGC_{optim}$ of the N+1th packet. For example,  if the sliding window is 10 then it is the data from 10 consecutive packets that are gathered + the $AGC_{optim}$ of the $11^th$. A collection of signals from internal metrics (LQI, SNR, CRC) is attached to a single wanted packet (/ interferer).

\subsection{ML model generation by training sequence}
\begin{figure}[!ht]
 \centering
      \includegraphics[scale=0.62]{ 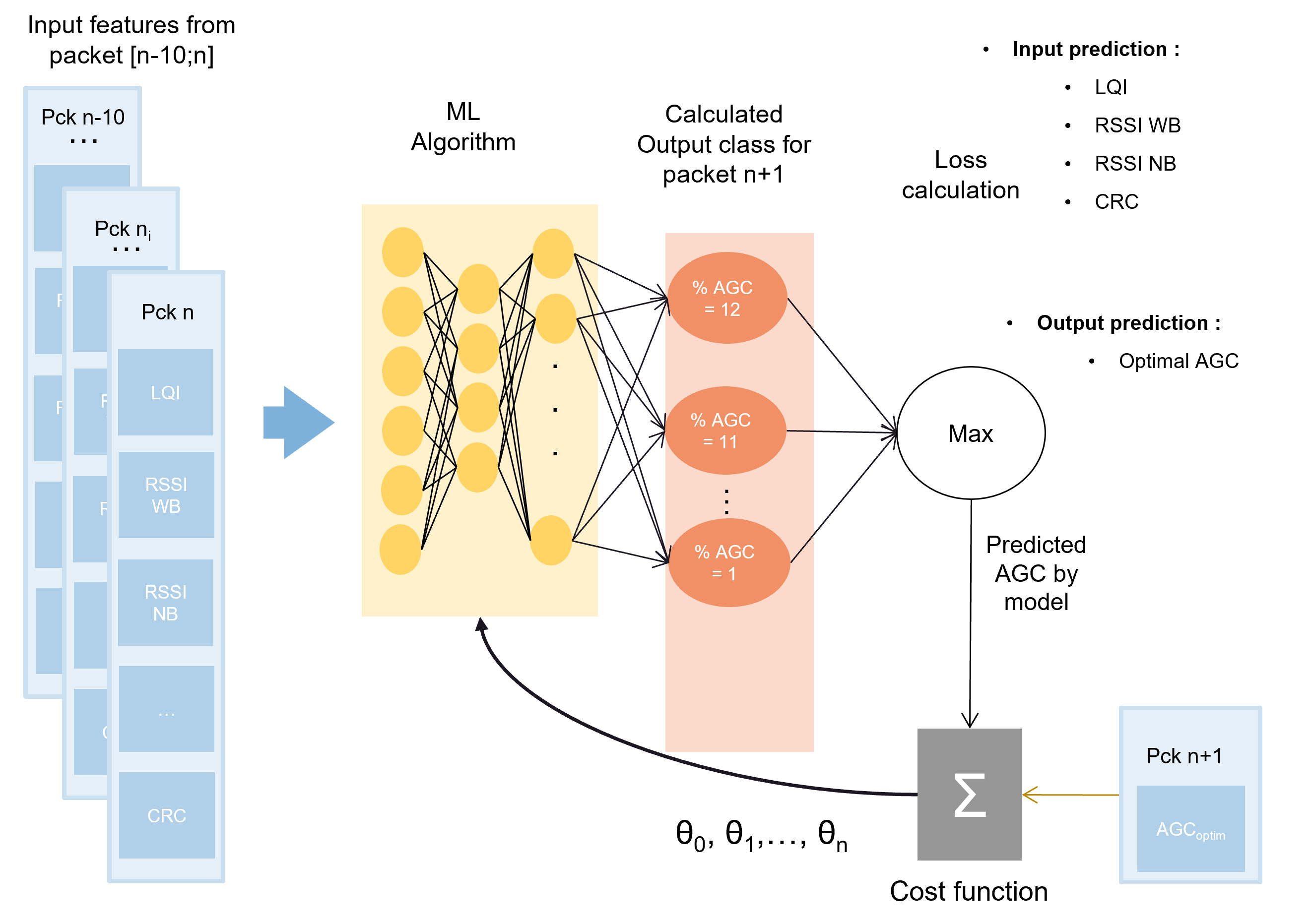}
    	\caption{ML model generation by training sequence}
        \label{fig:ML_scheme}
\end{figure}

During the training phase illustrated in Figure~\ref{fig:ML_scheme}, the data feeds the ML algorithm
\begin{itemize}
    \item ML internal coefficients are adjusted in steps until the ML model provides a result equal to the target $AGC_{optim}$ for Packet N+1 
\end{itemize}

The adaptation is done thanks to the loss function that compares the ML result and the target then it passes on the error to correct the coefficients afterwards. This process is done with all the data from the training set so the model can correct itself iteratively.

The test set is used to verify the accuracy of the whole model on different data and confirm that the dataset can generalise its process on similar but different data to avoid overfitting (symptom of overtrained model) and underfitting (symptom of undertrained model).

\newpage
\section{ML aided AGC in functional mode}
\subsection{Flowchart of ML data processing for AGC gain selection}

\begin{figure}[!ht]
 \centering
      \includegraphics[scale=0.62]{ 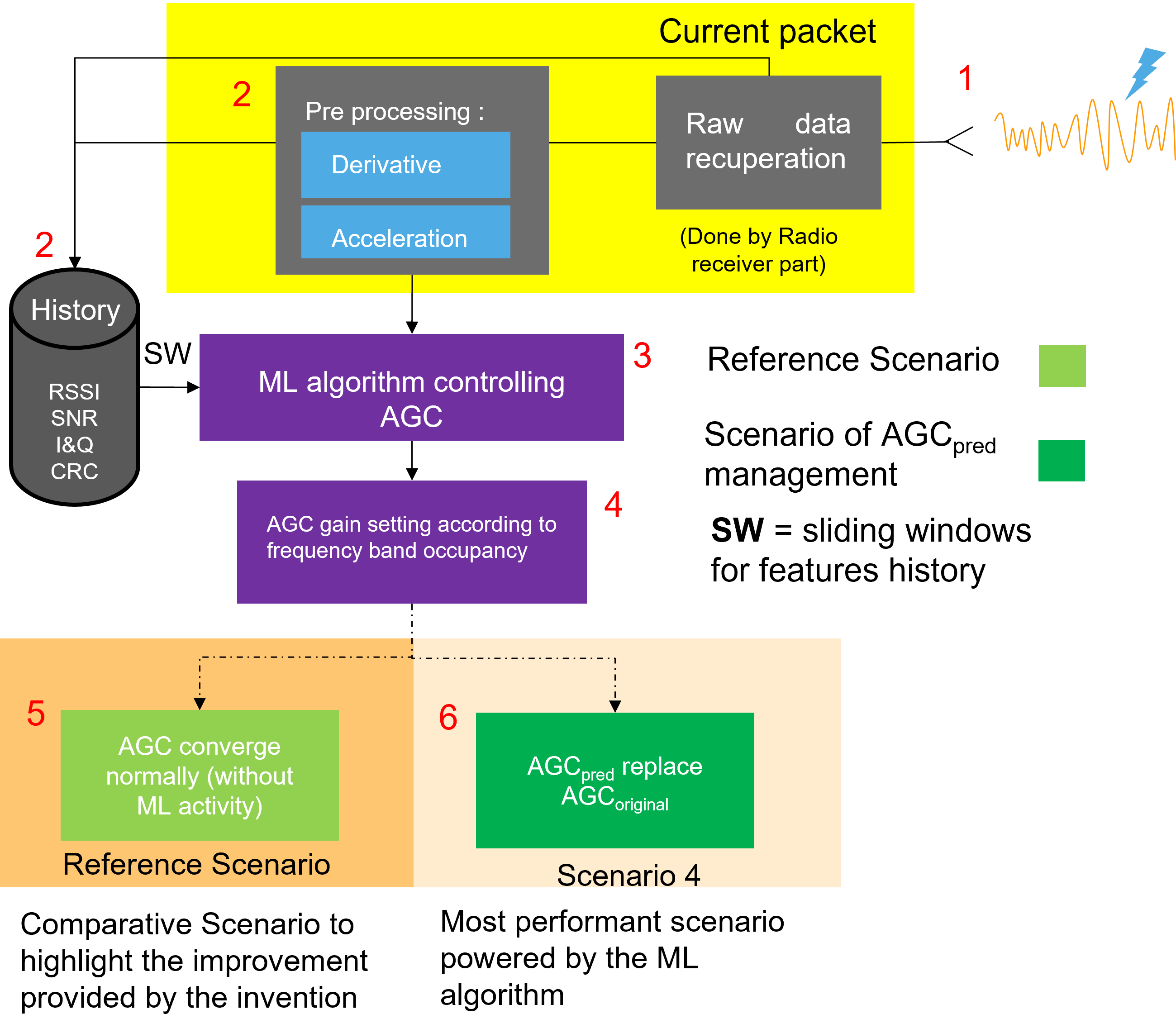}
      \caption{Flowchart of ML data processing for AGC gain selection}
        \label{fig:ML_processing}
\end{figure}

Once the ML model is trained, it is integrated into the receiver operating in normal mode.
ML operating in functional mode operates in the receiver as follows, with reference to Figure~\ref{fig:ML_processing}.
\begin{enumerate}

    \item At receiver warm-up then native/legacy AGC gain range is set with the upper limit ($AGC_{optim}$) as defined by ML at end of previous packet reception slot. During the current packet reception then raw data from receiver internal metrics are collected. So, the native/legacy AGC converges naturally without ML assistance.
    \item Then metrics are processed and stored in a buffer. For example, the metrics of the last 10 packets may be stored.
    \item Those metrics are used to predict with the ML algorithm the AGC index for the future slot of reception.
    \item This AGC can be used according to different scenarios, here are presented the reference scenario and the scenario (named “scenario 4”) where the predicted AGC index replaces the AGC index determined by legacy/native AGC. 
    \item For comparison during development, the reference scenario shows the performance of the radio with the native AGC algorithm with no ML feature
    \begin{itemize}
        \item Evaluate performance of the receiver
        \item Comparison point for the other scenario
    \end{itemize}

    \item The 4th scenario shows the performance of the radio when the AGC index (upper limit) predicted is used instead of the native AGC index
\end{enumerate}
\newpage
\subsection{Metrics (data) value collection during packet reception / sliding window}
\begin{figure}[!ht]
 \centering
      \includegraphics[scale=0.8]{ 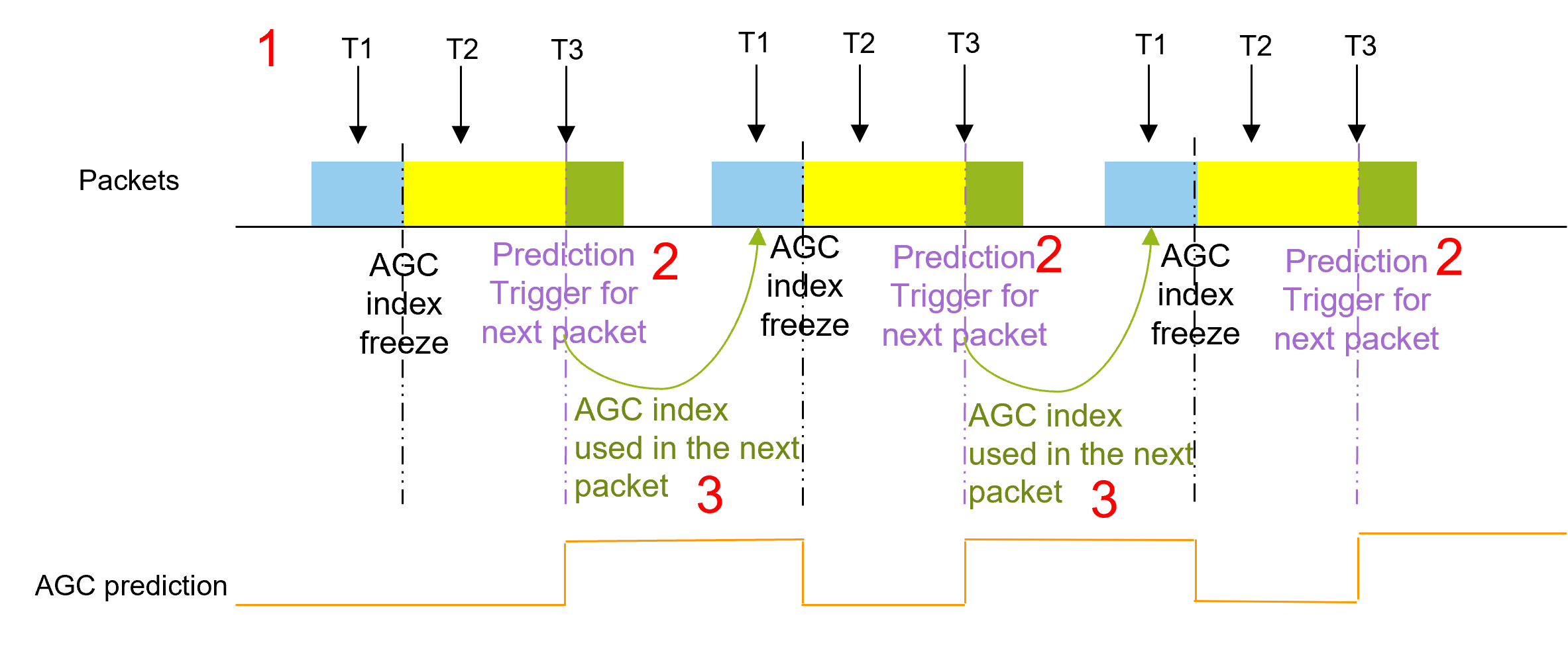}
    	\caption{Metrics (data) value collection during packet reception / sliding window}
        \label{fig:SLW}
\end{figure}

Metrics collection into the receiver during packet reception is performed as follow, with reference to Figure~\ref{fig:SLW}.

\begin{enumerate}
    \item During each packet reception, metrics value are collected at three timing location (T1,T2,T3) referenced to Sync signal assertion event (i.e. Access Address detection for BLE packet type) or expected start of packet
 \item T3 triggers the next packet AGC prediction by the ML model
 \item Prediction is used for next packet reception as initial  AGC index 
\begin{itemize}
    \item The AGC index freeze event triggers the reinitialization of the current metrics data buffer
\end{itemize}

\end{enumerate}

\subsection{Potential integration of the solution in Radio receiver (Scenario 4)}

\begin{figure}[!ht]
 \centering
      \includegraphics[scale=0.30]{ 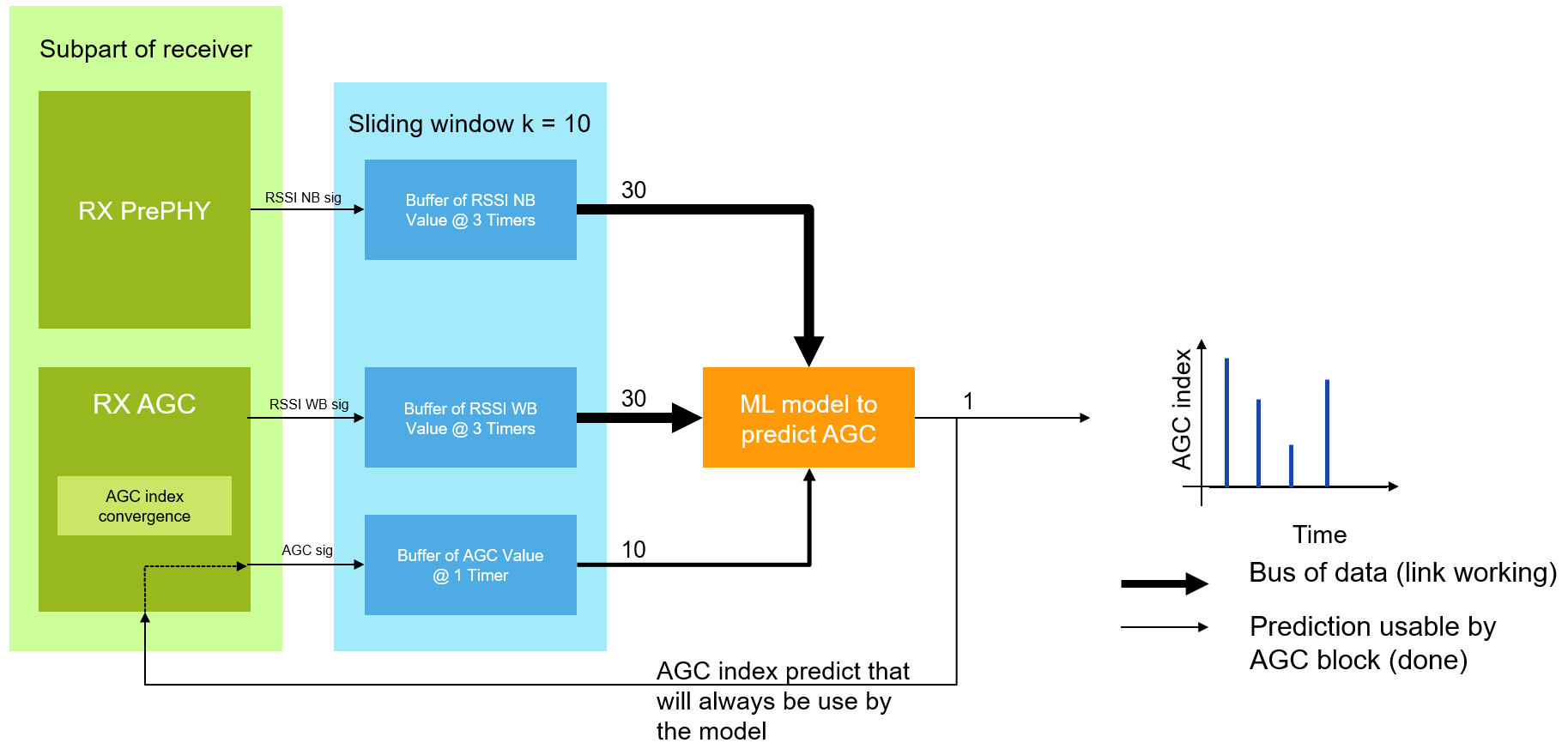}
    	\caption{Potential integration of the solution in Radio receiver}
        \label{fig:collection_data_RT}
\end{figure}

The Flowchart of the ML model integrated to the receiver which was modeled with Simulink is shown in Figure~\ref{fig:collection_data_RT}. The NB/WB RSSI metrics only represented for sake of simplification. Other metrics are used in practice. Input (metrics) of ML are collected in different parts of the Receiver. The ML model predicts the AGC index and replaces the original AGC index upper limit during the next wake up time slot.

\newpage

\subsection{ML usage by radio in functional mode (triggered at end of Nth packet reception)}
\begin{figure}[!ht]
 \centering
      \includegraphics[scale=0.55]{ 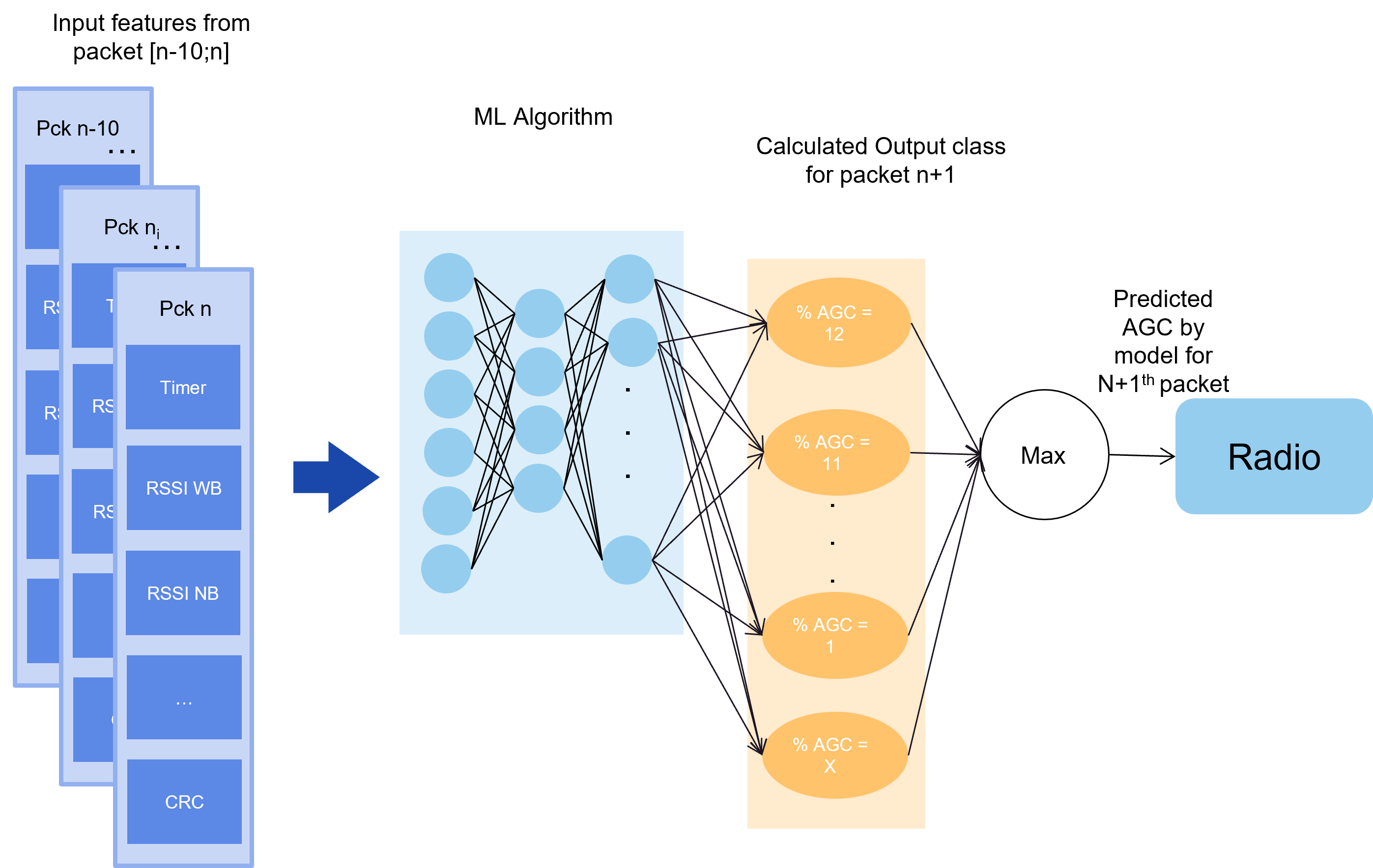}
    	\caption{ML usage by radio in functional mode (triggered at end of Nth packet reception)}
        \label{fig:ML_usage}
\end{figure}

In functional mode the AGC predicted by the ML model as illustrated in Figure~\ref{fig:ML_usage},  is directly used by the radio during the next wake up slot according to the scenario selected.

\subsection{Experiment result : PER versus power level of interferer located at 12 MHz frequency offset – Scenario 4}
\begin{figure}[!ht]
 \centering
      \includegraphics[scale=0.5]{ 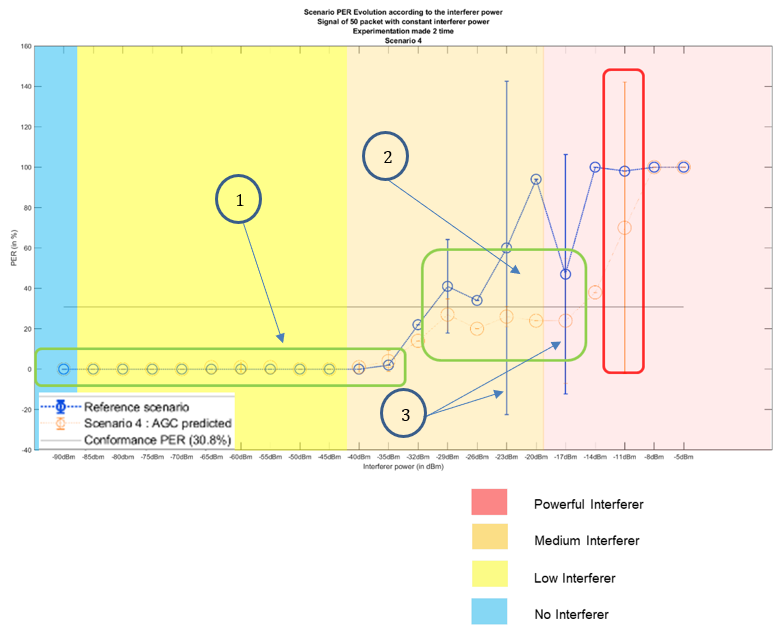}
    	\caption{Experiment result : PER versus interferer power level}
        \label{fig:scenario4}
\end{figure}

Figure~\ref{fig:scenario4}, shows the results of Packet Error Rate (PER) evolution versus interferer power level and versus AGC/ML operating mode.
AGC Reference scenario corresponds to native AGC free running operation (not assisted by ML).
AGC scenario 4 corresponds to ML-aided AGC operation .
The experiment settings condition  are :
\begin{itemize}
    \item BLE1Mbps BLE desired packet type
    \item Wanted power level : -60 dBm
    \item 50 desired signal packets
    \item Wi-Fi behavior : No pattern (continuous emission, not packetized)
    \item Blocker power level : Fixed power blocker according to the sweep range
    \item Scenario 4 :$AGC_{pred(icted)}$ replaces $AGC_{original}$(from legacy free running AGC)
    \item There are 3 repetitions of the experimentation to check repeatably of the results

\end{itemize}
\newpage

\subsection{Results summary}

Scenario 4 \& Reference scenario are on-par for interferer level below -35dBm – (1) 

Scenario 4 provides better performances than Reference scenario under the following conditions :
\begin{itemize}
    \item The Packet Error Rate (PER) is below standard requirement (30.8\%) for interferer level in between -29 dBm and -17 dBm power level range -  (2)
    \item No more native AGC instability at -29dBm, -23dBm \& -17dBm interferer levels (signal close to AGC gain switching) - (3)
\end{itemize}

\section{Conclusion}
This paper describes a new way of computing an AGC index. The statistical tool is based on a lightweight machine learning algorithm, integrated to the PHY layer.

The algorithm uses receiver metrics collected during previous receptions to analyse and anticipate the presence of interferer. This uses a reduced amount of data to optimise the computing.

The method uses determination (by the ML model ) and storage of optimum AGC gain setting at regular time points (at the end of the current packet reception) over a sliding time window when receiver in data communication and allows determination of a restricted gain range, based upon data feeding the ML algorithm, which is applied during the preamble phase of the next packet to be received.

The receiver can then extract the most appropriate gain level against past interferers and apply this as input to the gain adjustment before the gain freeze period (data payload reception).

The ML model using past information (from sliding window) allows capturing the most likely worst-case interferer and makes the system robust against interferers which may occur during the data payload reception.

The sliding time window allows adapting the receiver to the actual environmental conditions and to continuously configuring it to the highest sensitivity level whilst being robust against interferers.

\
\bibliographystyle{acm}
\bibliography{\jobname}

\end{document}